\documentclass[11pt,a4paper]{article}
\usepackage[T1]{fontenc}
\usepackage[utf8]{inputenc}
\usepackage{lmodern}
\usepackage[margin=25mm,headheight=14pt]{geometry}
\usepackage{amsmath,amssymb,bm}
\usepackage{graphicx,booktabs,tabularx,array,microtype,xcolor}
\usepackage{tikz,pgfplots}
\usetikzlibrary{arrows.meta,positioning,calc}
\pgfplotsset{compat=1.18}
\usepackage{caption,listings}
\usepackage[numbers,sort&compress]{natbib}
\usepackage{fancyhdr,hyperref,url}
\definecolor{ink}{HTML}{17344A}
\definecolor{teal}{HTML}{007F85}
\definecolor{pale}{HTML}{F0F5F7}
\hypersetup{colorlinks=true,linkcolor=ink,citecolor=ink,urlcolor=teal,
 pdftitle={Cavity-QED Molecular Dynamics: An eT--i-PI Interface},
 pdfauthor={Giuseppe Cassone},
 pdfsubject={Cavity Born--Oppenheimer dynamics and joint nuclear--photon path integrals}}
\newcommand{\Tr}{\operatorname{Tr}}
\newcommand{\ecbo}{E_{\mathrm{CBO}}}
\newcommand{\hf}{E_{\mathrm{HF}}}
\newcommand{\dd}{\mathrm{d}}
\newcommand{\code}[1]{\texttt{#1}}
\newcommand{\doi}[1]{\href{https://doi.org/#1}{doi:#1}}
\newcommand{\release}{2026-10-30-v4}

\title{\vspace{-7mm}\code{Cavity-QED Molecular Dynamics}:\\[3pt]
 An \code{eT}--\code{i-PI} Interface for \emph{Ab Initio} Cavity Born-Oppenheimer Dynamics and Joint Nuclear-Photon Path-Integral Sampling}
\author{Giuseppe Cassone\\[4pt]
 \small Institute for Chemical-Physical Processes, National Research Council of Italy (CNR-IPCF)\\
 \small Viale F. Stagno d'Alcontres 37, 98158 Messina, Italy\\[2pt]
 \small \href{mailto:giuseppe.cassone@cnr.it}{giuseppe.cassone@cnr.it}}
\date{\today}
\begin{document}
\maketitle
\begin{abstract}
Simulating molecules in quantum electrodynamic (QED) cavities requires a consistent description of electronic response, nuclear motion, and electromagnetic-field fluctuations. Here, I present \code{Cavity-QED Molecular Dynamics}, an interface connecting \code{eT} and \code{i-PI} for atomistic simulations of finite molecular systems coupled to a single photon mode. The code evaluates a self-consistent cavity Born-Oppenheimer Hartree-Fock (CBO-HF) energy surface and supports classical-nuclei dynamics under canonical (NVT) and microcanonical (NVE) ensembles, as well as canonical joint nuclear-photon path-integral sampling including nuclear quantum effects (NQEs). A fixed-displacement gradient correction provides nuclear and photon forces consistent with the same energy surface as the photon coordinate evolves independently. OpenMP threading, concurrent bead evaluations, and analysis utilities support practical simulations. Water-monomer benchmarks assess both dynamical propagation and equilibrium sampling. Energy-derivative checks and NVE timestep tests demonstrate consistent force evaluation and integration, while path integral molecular dynamics (PIMD) simulations with PILE-L thermostatting reproduce analytic finite-bead photon second moments, including their centroid and internal contributions, within sampling uncertainties. At the tested parameters, cavity coupling produces a small but statistically resolved increase in photon fluctuations, while changes in mean molecular geometry remain within sampling uncertainty. The resulting \code{Cavity-QED Molecular Dynamics} framework enables controlled investigations of how coupled quantum fluctuations of matter and the electromagnetic field influence molecular equilibrium properties within the CBO-HF approximation, providing a foundation for future studies of cavity-modified hydrogen bonding, vibrational spectra, and molecular polaritonic phenomena.
\end{abstract}

\section{Introduction}
Molecular cavity quantum electrodynamics (cavity-QED) connects electronic structure, vibrational motion, and electromagnetic degrees of freedom in a common Hamiltonian. For atomistic simulation, an essential question is how to describe the electronic response to a cavity coordinate while retaining the ability to sample anharmonic nuclear configurations. The cavity Born-Oppenheimer (CBO) approximation provides one route: electrons adjust to a specified nuclear geometry and photon displacement, and the resulting electronic energy defines a potential surface for the remaining coordinates.\cite{flick2017} Self-consistent CBO-HF formulations and their analytic gradients have already been developed for molecular ensembles and vibro-polaritonic spectra.\cite{schnappinger2023hf,schnappinger2023spectra,schnappinger2024correction} At each nuclear geometry and photon displacement, the self-consistent CBO-HF calculation reoptimizes the electronic density, accounting for cavity-induced electronic polarization. This response is absent from models that couple a field-free molecular dipole to a photon oscillator.\cite{fiechter2024} Related wavefunction-based developments include the coherent-state QED-HF formulation and its extension to correlated electron-photon states through QED coupled-cluster theory.\cite{Haugland2020}

There is also an established tradition of cavity molecular dynamics and path-integral treatment of coupled nuclear and photonic coordinates. Cavity molecular dynamics calculations have described classical liquid water \cite{li2020cavmd} and joint nuclear-photon quantum effects,\cite{li2022pimd} and \code{i-PI} includes a dedicated cavity force evaluator.\cite{ipifeatures} At the electronic-structure level, the recent \code{psi4-cbo} preprint provides another implementation of CBO calculations and derivatives.\cite{psi4cbo2026} Li, Car, and Flick recently introduced \code{cboamd}, a framework for machine-learning-driven simulations of molecular and collective vibrational strong coupling that combines zero-field energies, dipoles, and polarizabilities within an effective-field formulation of CBO theory.\cite{Li2026cboamd} The present work contributes a \code{eT}--\code{i-PI} interface realization that combines self-consistent electronic forces, a fixed-displacement gradient correction, explicit coordinate conventions, and reproducible verification for classical-nuclei molecular dynamics and joint path-integral sampling allowing for the simulation of nuclear quantum effects (NQEs) in \emph{ab initio} cavity-QED.

The interface separates potential evaluation from propagation and sampling. \code{eT} supplies the electronic calculation,\cite{et2020,et2026} and \code{i-PI} supplies the simulation engine taking care of the nuclear dynamics and sampling.\cite{ipi2024} One extra scalar coordinate represents the cavity boson mode. The same electronic model can therefore be used for one-bead trajectories or for a ring polymer containing both the nuclei and the photon coordinate. Because \code{i-PI} propagates the photon coordinate as an independent variable, nuclear forces must be calculated by differentiating the CBO energy with respect to nuclear positions while holding the photon displacement fixed. A nuclear-gradient expression that assumes an energy-minimized photon displacement can therefore give incorrect nuclear forces, even when the electronic energy and photon force are correct. The \code{eT} source snapshot used to build this interface originally employed a nuclear-gradient expression specialized to an optimized photon displacement. The present patch corrects this implementation by restoring the dipole self-energy and overlap contributions required at fixed photon displacement, yielding nuclear and photon forces consistent with the same CBO-HF energy surface.

This article documents release of the \code{cavity-QED Molecular Dynamics} software in its \code{\release} version while further developments and updates will be delivered in the future. Section~\ref{sec:theory} specifies the Hamiltonian and the finite-basis CBO-HF energy expression used in the implementation. Section~\ref{sec:implementation} describes the interface, followed by classical and path-integral sampling in Section~\ref{sec:sampling}. Section~\ref{sec:validation} presents numerical checks while Section~\ref{sec:molecular_sampling_benchmarks} presents several tests conducted on classical-nuclei and quantum nuclear-photon coupling molecular dynamics simulations of the water monomer. Section~\ref{sec:post-proc} describes analysis tools whereas the final Section is devoted to concluding remarks.

\section{Theory and implemented energy}\label{sec:theory}
\subsection{Single-mode Pauli-Fierz model and the CBO surface}
Atomic units are used unless stated otherwise. Let $\mathbf R$ denote all nuclear coordinates, $\mathbf r$ the electronic coordinates, and $q$ a mass-weighted cavity displacement with conjugate momentum $p_q$. For a single lossless mode in the long-wavelength, length-gauge description, I use
\begin{equation}
 \hat H=\hat T_{\mathrm n}+\hat H_{\mathrm{el}}(\mathbf R)
 +\frac{\hat p_q^2}{2}
 +\frac12(\omega\hat q+\lambda\hat\mu_\parallel)^2 ,
 \label{eq:pf}
\end{equation}
where $\omega$ is the angular frequency, $\lambda\mathbf e$ is the coupling vector, $\mathbf e$ is a unit polarization vector, and
\begin{equation}
 \hat\mu_\parallel=\mathbf e\cdot\left(\sum_A Z_A\mathbf R_A-\sum_i\mathbf r_i\right).
 \label{eq:dipole}
\end{equation}
$\hat H_{\mathrm{el}}$ includes nuclear repulsion. The plus sign in Eq.~\eqref{eq:pf} fixes the photon-force convention below; reversing $q$ gives the equivalent alternative convention. No rotating-wave or few-level truncation is introduced.

The cavity-dependent potential is
\begin{equation}
 \hat V_{\mathrm{cav}}=\frac12\omega^2q^2+\omega\lambda q\hat\mu_\parallel
 +\frac12\lambda^2\hat\mu_\parallel^2 .
 \label{eq:cavity}
\end{equation}
The last term is the dipole self-energy (DSE), retained together with the bilinear interaction. The ground-state electronic CBO surface is
\begin{equation}
 \ecbo(\mathbf R,q)=
 \min_{\Psi_{\mathrm{el}}}
 \langle\Psi_{\mathrm{el}}|\hat H_{\mathrm{el}}+\hat V_{\mathrm{cav}}
 |\Psi_{\mathrm{el}}\rangle .
 \label{eq:cbo}
\end{equation}
The implementation restricts this minimization to a normalized closed-shell Hartree-Fock determinant. Propagation or quantization of $(\mathbf R,q)$ then uses $\ecbo$ as its potential. Electronic nonadiabatic couplings and diagonal Born-Oppenheimer corrections are omitted in the current implementation. This way, the latter represents a ground-state surface approximation, rather than an exact solution of the electron-nuclear-photon problem.

\subsection{Finite-basis CBO-HF functional}
For reproducibility, the matrix form used by the released \code{eT} snapshot has to be explicitly specified. Let $S$ be the atomic-orbital (AO) overlap matrix, $D=C_{\mathrm{occ}}C_{\mathrm{occ}}^{\mathsf T}$ the occupied spatial-orbital density, and $P=2D$ the spin-summed RHF density. Thus $DSD=D$ and $\Tr(PS)=N_{\mathrm e}$. Define
\begin{equation}
 \widetilde\mu_{ab}=-\langle\chi_a|\mathbf e\cdot\mathbf r|\chi_b\rangle
 +\frac{\mu_{\mathrm n}}{N_{\mathrm e}}S_{ab},
 \qquad
 \mu_{\mathrm n}=\mathbf e\cdot\sum_A Z_A\mathbf R_A ,
 \label{eq:dipolematrix}
\end{equation}
and $d=\lambda\widetilde\mu$, $M=\Tr(Pd)=\lambda\langle\mu_\parallel\rangle$. Distributing the nuclear dipole over the electronic identity in Eq.~\eqref{eq:dipolematrix} recovers the total dipole upon taking its expectation. The implemented energy can be written
\begin{align}
 \ecbo&=\hf[D;\mathbf R]+\frac12\omega^2q^2+\omega qM+E_{\mathrm{DSE}},\label{eq:matrixenergy}\\
 E_{\mathrm{DSE}}&=\Tr(DdS^{-1}d)-\Tr(DdDd)+\frac12M^2 .
 \label{eq:dse}
\end{align}
Here $\hf$ is the ordinary molecular RHF energy evaluated with the cavity-relaxed density, including nuclear repulsion. The first two terms of Eq.~\eqref{eq:dse} express the dipole-fluctuation contribution in this representation; the last is the dipole Hartree term. Replacing the DSE by $\tfrac12M^2$ would change the model.

The one-electron DSE contribution in this snapshot uses the projected product $dS^{-1}d$.\cite{Lexander2025} In a finite AO basis, that product is generally different from an unprojected squared-position operator evaluated with quadrupole integrals. The independent check below reproduces the \emph{same projected convention}. Agreement tests implementation consistency, rather than equivalence of these finite-basis prescriptions or complete-basis convergence. This distinction matters when comparing different electronic-structure programs.

\subsection{Mapping the propagated coordinate to \code{eT}}
The developed interface uses \code{eT}'s fixed real coherent-displacement input, $z$, as a parameterization of $q$:
\begin{equation}
 z=q\sqrt{\frac{\omega}{2}}.
 \label{eq:mapping}
\end{equation}
Then $\omega z^2=\tfrac12\omega^2q^2$ and $\sqrt{2\omega}\,zM=\omega qM$. The mapping does not minimize the photon coordinate: \code{i-PI} supplies $q$, and \code{eT} relaxes the electrons at that specified value.

The returned potential follows Eq.~\eqref{eq:matrixenergy}, without an additional free-oscillator zero-point constant. A coordinate-independent constant has no effect on forces. Quantizing $q$ on this potential generates the oscillator zero-point energy through its kinetic energy and wavefunction, or through the path integral; the socket client does not add it separately.

\subsection{Forces at fixed photon displacement}
The required forces are
\begin{equation}
 F_{A\alpha}=-\left.\frac{\partial\ecbo}{\partial R_{A\alpha}}\right|_q,
 \qquad F_q=-\left.\frac{\partial\ecbo}{\partial q}\right|_{\mathbf R}.
 \label{eq:forces}
\end{equation}
At electronic stationarity the photon force is
\begin{equation}
 F_q=-\omega^2q-\omega M
 =-\omega^2q-\omega\lambda\langle\mu_\parallel\rangle .
 \label{eq:photonforce}
\end{equation}
AO functions do not depend on $q$, so this derivative has no nuclear-basis Pulay term. The \emph{total} dipole, including nuclei, is required. Nuclear forces also require ordinary HF integral derivatives, basis-overlap derivatives, and derivatives of all cavity terms.\cite{Lexander2025}

The distinction between a fixed and an optimized photon displacement is essential. At a stationary photon coordinate,
\begin{equation}
 q_*=-M/\omega ,
 \label{eq:qopt}
\end{equation}
where $M$ is evaluated self-consistently. At specified density, substitution cancels the mean-dipole contribution:
\begin{equation}
 \tfrac12\omega^2q_*^2+\omega q_*M+\tfrac12M^2=0 .
 \label{eq:cancellation}
\end{equation}
A gradient simplified using this cancellation cannot be used unchanged at arbitrary fixed $q$.

For example, at fixed density, differentiating the bilinear and dipole Hartree terms with respect to a nuclear coordinate $X$ gives
\begin{equation}
 (\omega q+M)\Tr(Pd_X),\qquad d_X=\partial d/\partial X .
 \label{eq:missing}
\end{equation}
In \code{eT}'s convention $g=\sqrt{\omega/2}\,d$, the contraction becomes
\begin{equation}
 \left(2z+\sqrt{\frac{2}{\omega}}M\right)\Tr(Pg_X).
 \label{eq:prefactor}
\end{equation}
The full $g_X$ includes the direct nuclear-dipole derivative. This term supplements, rather than replaces, the other integral and overlap derivatives. At $q=0$, $M$ generally remains nonzero: zero photon displacement does not remove the need for the DSE derivative. For the distributed \code{eT} implementation, the \code{v4} patch of the code restores these fixed-displacement terms and uses the actual harmonic overlap coefficient $\omega z^2/N_{\mathrm e}$. The overlap term proportional to $M^2$ associated with optimized bosons is retained only in the optimized branch. The energy and photon force are unchanged. 

\section{\code{Cavity-QED Molecular Dynamics} software architecture}\label{sec:implementation}
\subsection{An ordinary \code{i-PI} force client}
\begin{figure}[tb]
\includegraphics[width=\linewidth]{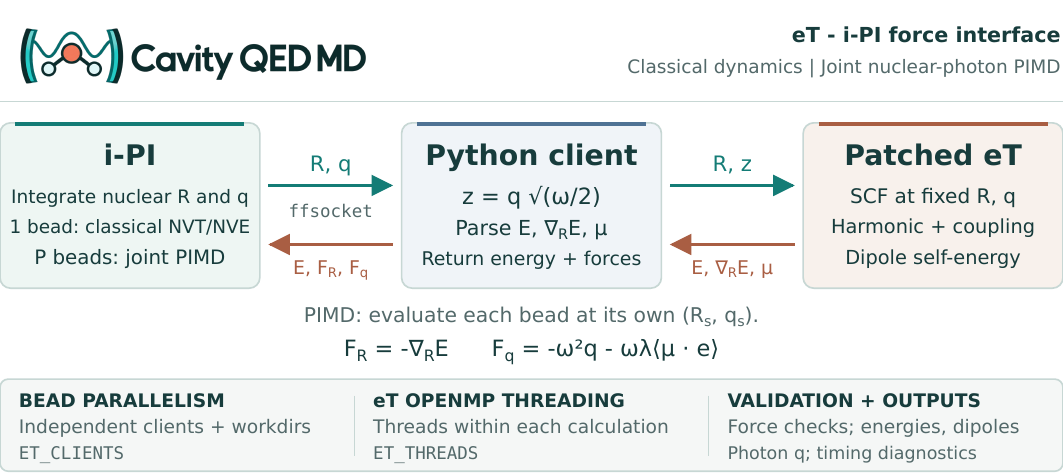}
\caption{Architecture and parallel execution of \code{Cavity-QED Molecular Dynamics}. \code{i-PI} propagates the nuclear coordinates $\mathbf R$ and photon coordinate $q$, or samples their joint ring-polymer distribution. The Python client maps $q$ to the coherent displacement $z=q\sqrt{\omega/2}$ and calls patched \code{eT} to obtain the energy, nuclear gradient, and total dipole. The energy includes the photon harmonic potential, light-matter coupling, and dipole self-energy. Teal arrows indicate configuration inputs and copper arrows indicate returned energies and derivatives. The client assembles consistent nuclear and photon forces and communicates with \code{i-PI} through ordinary \code{ffsocket}. Each PIMD bead is evaluated at its own ($\mathbf R_s,q_s$). Independent clients enable concurrent bead evaluations in separate working directories, while OpenMP threads parallelize each \code{eT} calculation.}
\label{fig:architecture}
\end{figure}

The Python client implements the unbatched base \code{i-PI} socket protocol (Fig.~\ref{fig:architecture}). At each request it receives coordinates, removes the photon pseudo-particle from the molecular geometry, generates a fixed-displacement \code{eT} input, and obtains the electronic energy, molecular gradient, and total dipole. It returns the negative nuclear gradient and Eq.~\eqref{eq:photonforce}, together with the energy, in atomic units.

The workflow uses ordinary \code{ffsocket}, with periodic boundary conditions disabled. It does not add cavity terms through \code{i-PI}'s separate \code{ffcavphsocket} evaluator. The \code{eT} energy already contains the photon harmonic potential, bilinear coupling and DSE; adding these terms again would double count them. At this stage, the molecular workflow does not supply a validated periodic stress or bulk NpT formulations, though work is planned also on this direction.

\subsection{Photon pseudo-particle and units}
The \code{i-PI} array contains all physical nuclei followed by a pseudo-particle labelled \code{L}. Its $x$ component stores $q$; the other two components are kept at zero. Its mass is one in electron-mass atomic units, consistent with $\hat p_q^2/2$. It is excluded from the \code{eT} geometry, nuclear charge, center of mass, and Eckart group. The storage axis for \code{L} is independent of physical polarization: storing $q$ in \code{L.x} does not imply laboratory-$x$ polarization. The client normalizes the separately supplied vector $\mathbf e$.

At the socket boundary all coordinates use internal atomic units. The workstation workflow's standalone XYZ input instead contains nuclear coordinates in {\AA}ngstrom ({\AA}) and \code{L.x} already in atomic units; its preparation utility converts only the nuclei. Embedded bead arrays and trajectories labelled \code{atomic\_unit} use atomic units throughout. The preserved PIMD XYZ initialization starts at $q=0$, avoiding a nonzero photon coordinate in its {\AA}-labelled initialization. Nonzero initialization must take into account this distinction.

\subsection{Restarts and parallel execution}
The workstation client can run the native \code{eT} executable or the standard launcher. It reuses the last converged SCF orbitals and retries a failed warm start from a superposition-of-atomic-densities guess. A restart is promoted only after a successful calculation; failed attempts are preserved. Rolling calculation directories limit routine per-step disk growth, with optional sparse snapshots. Structured records include energy, photon coordinate, dipole, timing, and restart status.

Parallel execution operates at two levels. Within each electronic-structure calculation, \code{eT} uses shared-memory OpenMP parallelism, with the thread count controlled by \code{ET\_THREADS}. Across ring-polymer beads, \code{i-PI} distributes independent force requests among multiple Python clients, whose number is set by \code{ET\_CLIENTS}. Each client evaluates the complete molecular system at the nuclear and photon coordinates of its assigned bead and maintains separate working directories and SCF restart files. Multiple bead calculations can therefore proceed concurrently without shared orbital-file writes. When fewer clients than beads are available, clients process successive requests until all required bead forces have been collected.

For $P_{\mathrm b}$ beads, $C$ clients and $T$ OpenMP threads per calculation, up to $\min(P_{\mathrm b},C)$ electronic evaluations can run simultaneously. Client and thread counts can be adjusted to balance bead concurrency against the parallel efficiency and memory requirements of individual calculations. The workstation launcher defaults to single-threaded BLAS and disables nested OpenMP parallelism to limit oversubscription; the supplied workflow does not require MPI. For one-bead dynamics, parallel acceleration comes primarily from threading within \code{eT}, whereas PIMD additionally exposes concurrency across beads. Utilities benchmark thread counts, compare execution paths and summarize timings, allowing the allocation to be selected for the target system and hardware.

The normal mode uses analytic nuclear gradients. An optional finite-difference mode evaluates nuclear derivatives from relaxed energies for checks or diagnosis. Central differences require two displaced energies per nuclear component in addition to the reference calculation.

\section{Classical dynamics and joint path integrals}\label{sec:sampling}
\subsection{One-bead dynamics and global Eckart constraints}
For a single bead the propagated Hamiltonian is
\begin{equation}
 H_{\mathrm{cl}}=\sum_A\frac{\mathbf P_A^2}{2M_A}
 +\frac{p_q^2}{2}+\ecbo(\mathbf R,q).
 \label{eq:classical}
\end{equation}
The workstation utility prepares ordinary NVT/NVE inputs for a water monomer or a supplied 21-water cluster. The photon is explicit in both cases. 

A separate one-bead NVE option removes nuclear center-of-mass translation and overall rotation using a global Eckart reference. For centered reference coordinates $\mathbf R_A^0$, the constraints have the form
\begin{equation}
 \sum_A M_A\mathbf R_A=0,\qquad
 \sum_A M_A\mathbf R_A^0\times\mathbf R_A=0 .
 \label{eq:eckart}
\end{equation}
Momenta are projected onto the compatible tangent space. The photon is excluded. In a cluster, the global constraint removes overall motion while retaining internal inter- and intramolecular motion. It restricts reorientation relative to the laboratory cavity polarization and therefore defines a constrained physical comparison. The released Eckart path is one-bead NVE; a compatibility wrapper adapts the pinned \code{i-PI} constrained-dynamics interfaces without modifying the upstream installation.

\subsection{Equilibrium quantization of nuclei and photon}
On the same surface the quantum slow-coordinate Hamiltonian is
\begin{equation}
 \hat H_{\mathrm{slow}}=-\sum_A\frac{\nabla_A^2}{2M_A}
 -\frac12\frac{\partial^2}{\partial q^2}+\ecbo(\mathbf R,q).
 \label{eq:slow}
\end{equation}
For $\beta=(k_{\mathrm B}T)^{-1}$ and $P_{\mathrm b}$ beads, its primitive imaginary-time discretization has configurational weight
\begin{align}
 \rho_{P_{\mathrm b}}&\propto\exp\left[-\frac{\beta}{P_{\mathrm b}}
 \sum_{s=1}^{P_{\mathrm b}}\{\ecbo(\mathbf R_s,q_s)+U_s^{\mathrm{spring}}\}\right],
 \label{eq:pimd}\\
 U_s^{\mathrm{spring}}&=\frac{\omega_{P_{\mathrm b}}^2}{2}
 \left[\sum_A M_A|\mathbf R_{A,s}-\mathbf R_{A,s+1}|^2+(q_s-q_{s+1})^2\right],
 \quad \omega_{P_{\mathrm b}}=\frac{P_{\mathrm b}}{\beta}.
 \label{eq:springs}
\end{align}
The bead index is cyclic. The photon spring and physical photon harmonic potential have different origins and are both required.

Each bead calls \code{eT} at its own $(\mathbf R_s,q_s)$. \code{i-PI} supplies ring-polymer integration and thermostatting. The preserved monomer input uses PILE-L,\cite{pile2010} eight beads at 300~K and 100 steps of 0.1~fs. Of course, these values represent demonstration settings only and not converged production prescriptions. A position-dependent observable is estimated by averaging over beads. In particular, $\langle q^2\rangle$ requires the bead-averaged square; the squared centroid omits internal ring-polymer fluctuations. As $P_{\mathrm b}\rightarrow\infty$, PIMD samples the equilibrium quantum statistics of Eq.~\eqref{eq:slow}, within the stated electronic/CBO approximations. Of course, a thermostatted PIMD trajectory does not represent a physical quantum real-time evolution. The equilibrium PIMD trajectories used here do not directly provide physical real-time correlation functions or infrared spectra. 

\subsection{Analytic photon benchmark for bead convergence}
\begin{figure}[tb]
\centering
\begin{tikzpicture}
\begin{axis}[width=0.86\linewidth,height=61mm,
 xmode=log,log basis x=2,ymode=log,
 xmin=1,xmax=145,ymin=0.15,ymax=110,
 xtick={1,4,8,16,32,64,128},xticklabels={1,4,8,16,32,64,128},
 xlabel={Number of primitive beads, $P_{\mathrm b}$},
 ylabel={Variance underestimate (\%)},
 grid=major,grid style={black!10},tick label style={font=\small},
 label style={font=\small},legend cell align=left,
 legend style={font=\small,at={(0.98,0.98)},anchor=north east,
 fill=white,draw=black!20}]
 \addplot[color=teal,thick,mark=*] coordinates {
 (1,88.594917)(4,58.494327)(8,32.598690)(16,12.304492)
 (32,3.5548589)(64,0.92544682)(128,0.23379106)};
 \addlegendentry{$\omega=0.01666$ a.u.}
 \addplot[color=ink,thick,mark=square*] coordinates {
 (1,89.866473)(4,62.434453)(8,37.025827)(16,14.886634)
 (32,4.4406503)(64,1.1679599)(128,0.29586805)};
 \addlegendentry{$\omega=0.0187505$ a.u.}
 \addplot[black!50,dashed,domain=1:145] {1};
 \node[font=\scriptsize,anchor=south west,fill=white,inner sep=1pt]
 at (axis cs:1.4,1) {1\% reference};
\end{axis}
\end{tikzpicture}
\caption{Analytic free-photon variance convergence at 300~K. Values are $100[1-\langle q^2\rangle_{P_{\mathrm b}}/\langle q^2\rangle_{\mathrm Q}]$ from Eqs.~\eqref{eq:variance} and \eqref{eq:pvariance}. This is a primitive-discretization reference, not a simulated trajectory.}
\label{fig:beads}
\end{figure}
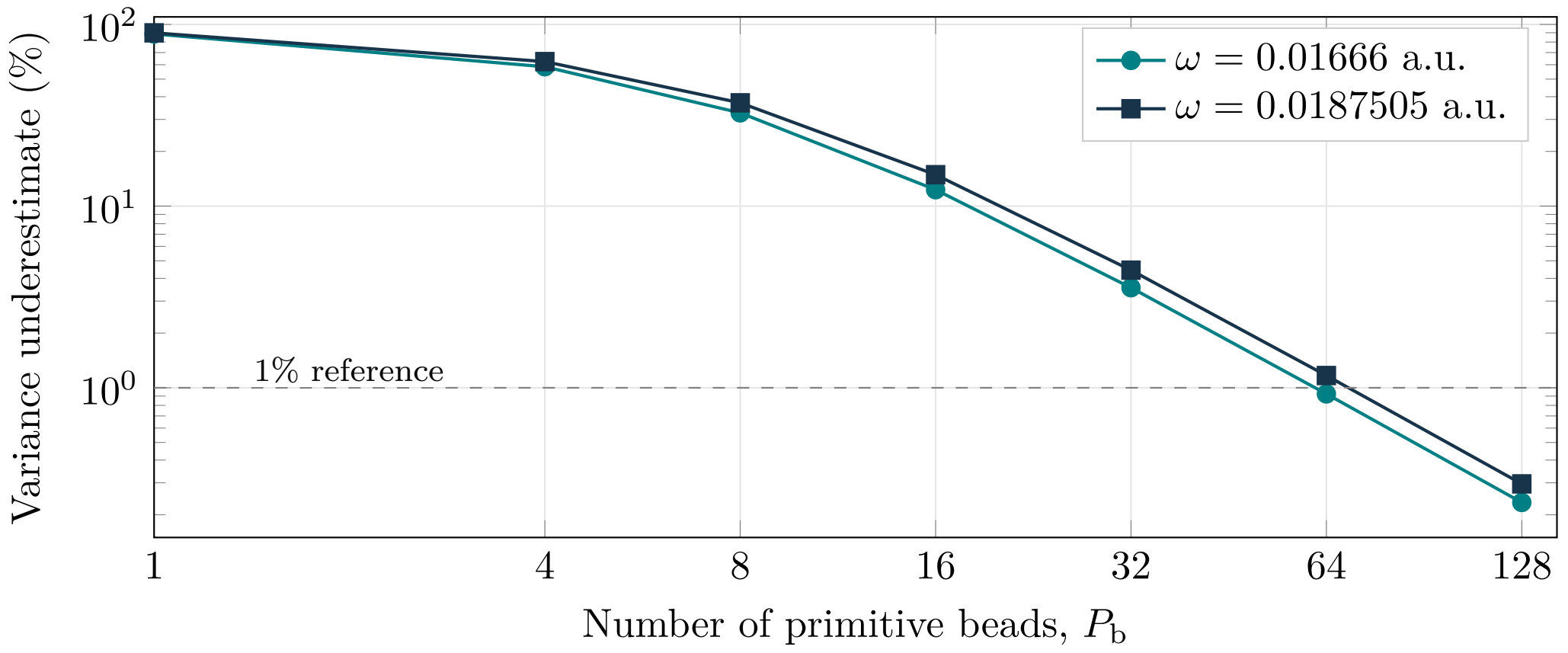

At $\lambda=0$, the photon separates from molecular coordinates, and
\begin{equation}
 \langle q^2\rangle_{\mathrm Q}=\frac{1}{2\omega}
 \coth\left(\frac{\beta\omega}{2}\right),
 \qquad \langle q^2\rangle_{\mathrm{cl}}=\frac{1}{\beta\omega^2}.
 \label{eq:variance}
\end{equation}
Diagonalizing the harmonic ring polymer gives the finite-bead primitive result independently of a trajectory:
\begin{equation}
 \langle q^2\rangle_{P_{\mathrm b}}=\frac1\beta
 \sum_{k=0}^{P_{\mathrm b}-1}
 \left[\omega^2+4\omega_{P_{\mathrm b}}^2
 \sin^2\left(\frac{\pi k}{P_{\mathrm b}}\right)\right]^{-1}.
 \label{eq:pvariance}
\end{equation}
Fig.~\ref{fig:beads} evaluates this expression for the distributed frequencies. At the PIMD default $\omega=0.01666$~a.u. and 300~K, $\beta\omega=17.54$ and the exact variance is $30.012$~a.u. Eight primitive beads give $20.228$~a.u., a 32.60\% underestimate; 64 beads reduce this error to 0.93\%. The one-bead variance is $3.423$~a.u. These estimates are analytic discretization values and do not correspond to explicit trajectory measurements. In other words, they simply provide a sensitive check of the photon mass, units, and bead treatment. 

\subsection{Interpreting nuclear and photonic quantum effects}
Nuclear and photonic quantum effects are replicated in the PIMD example supplied in the \code{cavity-QED Molecular Dynamics} code. Comparing them with one-bead dynamics therefore changes the statistical treatment of nuclei and photon simultaneously. At matched model parameters, the difference measures a joint classical-to-quantum difference, rather than a nuclear-only or photon-only correction. For an observable $A$, four consistently specified calculations could define
\begin{equation}
 \Delta A_{\mathrm{mix}}=
 A_{\mathrm{Qn,Qq}}-A_{\mathrm{Qn,Cq}}-A_{\mathrm{Cn,Qq}}+A_{\mathrm{Cn,Cq}},
 \label{eq:mixed}
\end{equation}
where Q and C indicate quantum and classical statistics. This represents a proposed comparison protocol only: all four mixed ensembles are not validated in the current release.

\section{Verification and numerical checks}\label{sec:validation}
\subsection{Independent molecular energy and derivative checks}
The molecular test uses the neutral water geometry supplied in the \code{cavity-QED Molecular Dynamics} code, cc-pVDZ,\cite{Dunning1989} laboratory-$z$ polarization and $\omega=0.0187505$~a.u. An independent PySCF 2.14.0 implementation\cite{pyscf2020} reconstructs Eqs.~\eqref{eq:matrixenergy}--\eqref{eq:dse} and relaxes the orbitals at every displaced geometry. The repository script \code{scripts/check\_fixed\_displacement\_gradient.py} evaluates all nine nuclear components for $\lambda=0$ and 0.01, each with $q=0$ and 0.5.

At $\lambda=0.01$, $q=0$, the independent energy is $-75.98957604860672$~hartree, compared with the reported \code{eT} value $-75.98957604860701$~hartree, a difference of approximately $2.9\times10^{-13}$~hartree. This agreement tightly checks the energy and finite-basis conventions.


The independent gradient combines ordinary PySCF HF derivatives with cavity terms whose overlap and dipole-matrix derivatives are obtained by central differences. It checks the fixed-displacement algebra against fully relaxed total-energy differences; it is not a second all-analytic integral implementation. Its SCF energy threshold is $10^{-13}$~hartree and its orbital-gradient threshold is $10^{-10}$. The distributed calculation was performed using Python 3.12.14, PySCF 2.14.0, NumPy 2.3.5, SciPy 1.17.0 and h5py 3.16.0. 

\subsection{Interface tests and constrained integration}
The regression suite exercises protocol messages, parsing, analytic/finite-difference agreement for a harmonic electronic test double, execution backends, SCF recovery, staging, and constrained-integration support. A separate integration test uses real \code{i-PI} with its \emph{internal harmonic force field}, rather than \code{eT}. Archived 30-step tests at 0.1~fs yield Table~\ref{tab:eckart}. They check constraints and exclusion of inactive photon directions. The classical IR utility also passes its synthetic-signal self-test for frequencies, center-of-mass correction, photon exclusion, molecular frames, and multi-trajectory output.

\begin{table}[tb]
\centering
\caption{Archived real-i-PI internal-harmonic Eckart checks over 30 steps. Residuals use i-PI's internal normalization and atomic units. Energy ranges come from printed conserved-energy output and are not electronic CBO-HF drift measurements.}
\label{tab:eckart}
\small
\begin{tabular}{lrrr}
\toprule
System & Position residual & Velocity residual & Energy range (eV) \\
\midrule
One water + $q$ & $3.35\times10^{-16}$ & $1.08\times10^{-20}$ & $3.05\times10^{-5}$ \\
21 waters + $q$ & $1.31\times10^{-14}$ & $7.25\times10^{-20}$ & $2.10\times10^{-6}$ \\
\bottomrule
\end{tabular}
\end{table}

\subsection{Target-machine validation and interpretation}
The latest \code{cavity-QED Molecular Dynamics} code distribution (\code{v4}) and documented workflow were reported as working on 30th of October 2026. On each new installation, \code{validate\_forces.py --all-nuclear} checks all nine monomer nuclear components and the photon component against relaxed-energy differences at the four $(\lambda,q)$ combinations above. Its default absolute tolerance is $10^{-5}$ atomic units for each tested component. The backend-comparison script checks native and launcher execution. Larger applications require checks on representative geometries and photon coordinates using their actual templates.

\section{Molecular dynamics and equilibrium sampling benchmarks}
\label{sec:molecular_sampling_benchmarks}

The molecular dynamics benchmarks here reported test the complete connection between the compiled \texttt{eT} electronic-structure calculation and the propagation and sampling performed by \texttt{i-PI}. Energy conservation is examined in one-bead microcanonical trajectories, while joint nuclear-photon path-integral simulations test equilibrium sampling against an analytic reference and
resolve a small coupling-dependent response. These complementary checks
connect the energy and force verification to explicit molecular trajectories.

\subsection{Molecular system and sampling protocol}
\label{subsec:benchmark_protocol}

The benchmark simulations here described use a single water molecule on the self-consistent CBO-HF surface, a cc-pVDZ basis, laboratory-$z$ polarization, and a photon frequency $\omega=0.0187505$~a.u. The two coupling strengths explored are
$\lambda=0$ and $0.01$~a.u. Nuclear and photon forces are supplied by the
analytic-force workflow. The global Eckart option capable of constraining indicated nuclear degrees of freedom is disabled for these benchmarks.

The equilibrium simulations use $P_{\mathrm b}=32$ beads for both the
nuclei and the photon coordinate, a timestep of $0.1$~fs, and a target
temperature of $300$~K. PILE-L thermostatting\cite{pile2010} is used with
a centroid thermostat time constant of $25$~fs. PILE-L was selected because it provided more reliable convergence of the photon estimators in the present benchmarks. Local thermostatting is a useful choice for the uncoupled benchmark because the photon is then dynamically separated from the molecule. Two replicas, initialized with seeds 11 and 29, are run at each coupling. Each replica contains $0.5$~ps of equilibration followed by $10$~ps of production sampling, with configurations recorded every $1$~fs. The same seed is used for matching coupled and uncoupled runs, allowing their statistical covariance to be retained in the comparison.

\subsection{Energy conservation and timestep dependence}
\label{subsec:benchmark_nve}

One-bead, classical-nuclei NVE trajectories provide a direct test of the energy returned by \texttt{eT} and the forces used by \texttt{i-PI} on the same CBO-HF
surface. Each coupling is tested over $2$~ps with timesteps of $0.10$ and
$0.05$~fs, using matched starting coordinates and momenta for the two
timesteps. The initial conditions serve as an energy-conservation stress
test. Table~\ref{tab:benchmark_nve} reports the range and standard deviation
of the total nuclear-photon energy, together with the slope of a linear
fit to its time dependence.

Halving the timestep reduces the energy standard deviation by factors of
$4.00$ at zero coupling and $4.08$ at $\lambda=0.01$~a.u. Thus, halving the timestep reduces the total-energy fluctuations by approximately a factor of four, as expected for a second-order integration scheme. The coupled energy range decreases from $2.28$ to $0.57$~meV, while the fitted drift at $0.05$~fs is $-1.73\times10^{-5}$~meV\,ps$^{-1}$. The largest absolute fitted drift among the four trajectories is $2.66\times10^{-3}$~meV\,ps$^{-1}$. Together with the independent energy-derivative checks, these results support consistent force-driven propagation by the compiled molecular workflow over the tested interval. The fitted slopes characterize these short $2$~ps trajectories and do not test eventual longer-time drift bounds.
\begin{table}[!htbp]
  \centering
  \caption{Energy conservation in $2$~ps-long trajectories of a classical-nuclei water monomer plus a single cavity mode in the NVE ensemble. Energy ranges and standard deviations (SDs) refer to the full system, without normalization per atom. The drift is the least-squares slope of the total energy versus time.}
  \label{tab:benchmark_nve}
  \vspace{0.4em}
  \small
  \begin{tabular}{ccrrr}
    \hline
    $\lambda$ (a.u.) & $\Delta t$ (fs) &
    \shortstack{Energy range\\(meV)} &
    \shortstack{Energy SD\\(meV)} &
    \shortstack{Fitted drift\\(meV\,ps$^{-1}$)} \\
    \hline
    $0$    & $0.10$ & $2.34$ & $0.505449$ & $ 2.659\times10^{-3}$ \\
    $0$    & $0.05$ & $0.59$ & $0.126308$ & $ 5.307\times10^{-4}$ \\
    $0.01$ & $0.10$ & $2.28$ & $0.481029$ & $ 1.264\times10^{-3}$ \\
    $0.01$ & $0.05$ & $0.57$ & $0.117779$ & $-1.730\times10^{-5}$ \\
    \hline
  \end{tabular}
\end{table}

\subsection{Sampling the uncoupled photon}
\label{subsec:benchmark_free_photon}

At $\lambda=0$, the photon behavior separates from the molecular coordinates and
provides a quantitative equilibrium sampling benchmark. For photon bead
coordinates $q_s$, define
\begin{equation}
  \bar q=\frac{1}{P_{\mathrm b}}\sum_{s=1}^{P_{\mathrm b}}q_s,
  \qquad
  Q_2=\frac{1}{P_{\mathrm b}}\sum_{s=1}^{P_{\mathrm b}}q_s^2,
  \qquad
  S_q=\frac{1}{P_{\mathrm b}}\sum_{s=1}^{P_{\mathrm b}}(q_s-\bar q)^2.
  \label{eq:benchmark_photon_estimators}
\end{equation}
The physical second moment is estimated as
$\langle q^2\rangle_{P_{\mathrm b}}=\langle Q_2\rangle$, and
$Q_2=\bar q^2+S_q$ for every configuration. Resolving these two contributions
therefore checks sampling more stringently than comparing the total
second moment alone.

For $\beta=(k_{\mathrm B}T)^{-1}$ and $\omega_{P_{\mathrm b}}=P_{\mathrm b}/\beta$,
the analytic harmonic references for the finite-bead primitive
configurational distribution are
\begin{align}
  \langle Q_2\rangle_{P_{\mathrm b}}
    &=\frac{1}{\beta}\sum_{k=0}^{P_{\mathrm b}-1}
      \left[\omega^2+4\omega_{P_{\mathrm b}}^2
      \sin^2\!\left(\frac{\pi k}{P_{\mathrm b}}\right)\right]^{-1},
      \label{eq:benchmark_finite_bead}\\
  \langle\bar q^2\rangle_{P_{\mathrm b}}
    &=\frac{1}{\beta\omega^2},
  \qquad
  \langle S_q\rangle_{P_{\mathrm b}}
    =\langle Q_2\rangle_{P_{\mathrm b}}-\frac{1}{\beta\omega^2}.
      \label{eq:benchmark_centroid_reference}
\end{align}
At the parameters used here, the total, centroid, and internal references
are $25.48181$, $2.70220$, and $22.77961$~a.u., respectively.

\begin{figure}[t]
  \centering
  \includegraphics[width=\linewidth]{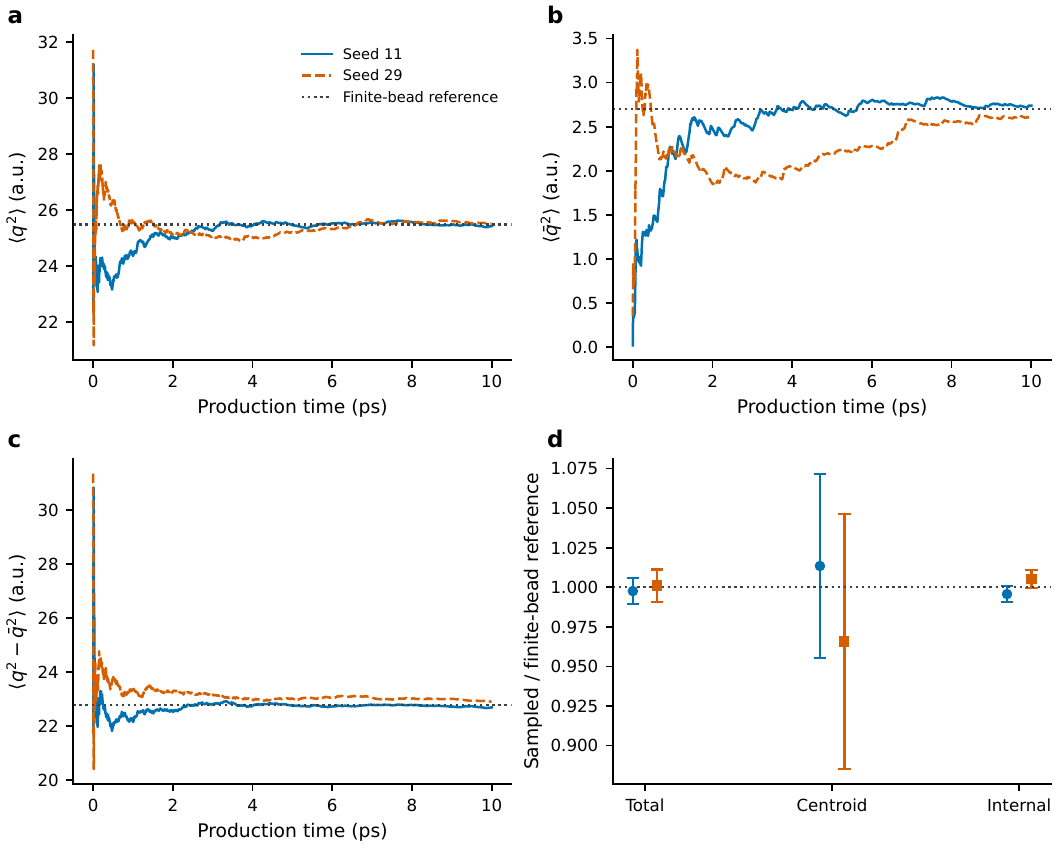}
  \caption{Trajectory-based validation of uncoupled photon sampling for
  a water monomer at $T=300$~K, $P_{\mathrm b}=32$,
  $\omega=0.0187505$~a.u., and $\Delta t=0.1$~fs, using PILE-L
  thermostatting. Cumulative production estimates of (a) the
  bead-averaged photon second moment $\langle Q_2\rangle$,
  (b) the squared centroid $\langle\bar q^2\rangle$, and
  (c) the internal spread $\langle S_q\rangle$ are shown for two
  replicas with different initialization seeds. Dotted horizontal lines give the finite-bead harmonic references. (d) Production averages divided by their respective references; error bars show one standard error from $500$~fs time
  blocks. Each replica involves $10$~ps of production after
  $0.5$~ps of equilibration. The horizontal axes in (a)-(c) measure
  PIMD sampling time.}
  \label{fig:benchmark_free_photon}
\end{figure}

Fig.~\ref{fig:benchmark_free_photon} shows the cumulative estimates of
these three quantities and their production averages. The two total
second-moment estimates are $25.420\pm0.206$ and $25.508\pm0.260$~a.u.,
where the uncertainties are one standard error from $500$~fs blocks.
The centroid estimates are $2.739\pm0.158$ and
$2.610\pm0.218$~a.u.; the corresponding internal contributions are
$22.682\pm0.121$ and $22.899\pm0.128$~a.u. All six estimates lie within
one estimated standard error of their references. In particular,
Fig.~\ref{fig:benchmark_free_photon}-d shows that the agreement of the
total is accompanied by agreement of its separate contributions.
The centroid exhibits slower cumulative fluctuations than the internal
spread, illustrating that its equilibration has to be assessed explicitly.

These trajectory measurements test the photon-coordinate conventions,
ring-polymer sampling, and estimator implementation against an
independently known equilibrium distribution. The comparison is made
with the finite-bead reference. This reference remains $4.44\%$ below
the exact quantum harmonic second moment,
$[2\omega]^{-1}\coth(\beta\omega/2)=26.66596$~a.u.

\subsection{Coupled equilibrium response}
\label{subsec:benchmark_coupled_response}

The same protocol at $\lambda=0.01$~a.u. is aimed at testing the joint nuclear-photon sampling on the coupled CBO-HF surface. The coupled photon second moments
are $25.493\pm0.203$ and $25.582\pm0.257$~a.u. for seeds 11 and 29,
respectively, with the same one-standard-error convention. These values
are mutually consistent and are close to the uncoupled results. Their
separate uncertainties, however, do not specify the uncertainty of a
matched-seed coupling difference.

For an observable $A$, the two conditions are subtracted at corresponding
production times within each seed pair. If $d_b$ is the resulting mean
difference in block $b$, the paired estimate and its standard error are
\begin{equation}
  \widehat{\Delta A}=\frac{1}{M}\sum_{b=1}^{M}d_b,
  \qquad
  \mathrm{SE}(\widehat{\Delta A})=
  \left[\frac{1}{M(M-1)}\sum_{b=1}^{M}
  (d_b-\widehat{\Delta A})^2\right]^{1/2}.
  \label{eq:benchmark_paired_error}
\end{equation}
Here, $\Delta$ denotes the $\lambda=0.01$ result subtracted by the $\lambda=0$
result, and $M=20$ contiguous $500$~fs blocks are used for each seed.
The final endpoint frame is omitted to retain complete equal blocks.
This procedure includes the covariance between separate conditions whilst individual beads are not counted as independent samples. Per-pair confidence
intervals use Student's $t$ distribution. The two independent seed-pair estimates are averaged and their uncertainties are combined to obtain a 95\% confidence interval.

\begin{figure}[!htbp]
  \centering
  \includegraphics[width=0.98\linewidth]{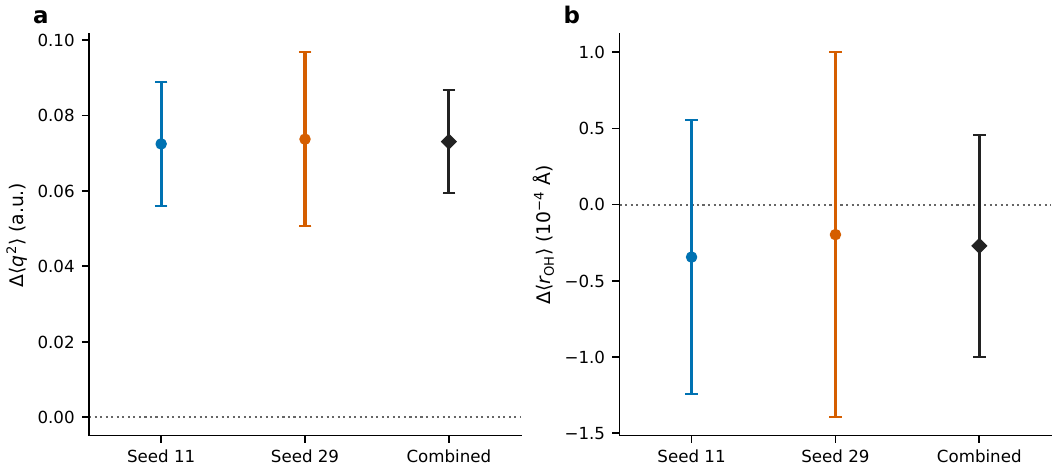}
  \caption{Paired equilibrium response to cavity coupling at
  $T=300$~K, $P_{\mathrm b}=32$, $\Delta t=0.1$~fs, and
  $\omega=0.0187505$~a.u. Differences are defined as the
  $\lambda=0.01$ result minus the $\lambda=0$ result.
  (a) Change of the photon second moment.
  (b) Change of the mean O--H distance, expressed in units of
  $10^{-4}$~\AA{}. Colored circles show the estimates from two
  matched-seed pairs; black diamonds show their equal-weight
  combination. Each condition and seed refers to $10$~ps of production
  sampling following $0.5$~ps of equilibration. Error bars are
  $95\%$ confidence intervals obtained from $500$~fs paired blocks,
  retaining the covariance between matching conditions. Dotted lines
  indicate zero change. These intervals quantify sampling uncertainty
  at the specified bead count and timestep.}
  \label{fig:benchmark_paired_response}
\end{figure}

The photon block means are strongly correlated between matching
conditions, with correlation coefficients of approximately $0.99940$
and $0.99915$ for the two seed pairs. Consequently, the paired difference
can be resolved much more precisely than the separate-condition
averages. Fig.~\ref{fig:benchmark_paired_response}-a gives a combined
increase
\begin{equation}
  \Delta\langle Q_2\rangle=0.0731~\mathrm{a.u.},
  \qquad 95\%\ \mathrm{CI}=[0.0593,\,0.0868]~\mathrm{a.u.},
  \label{eq:benchmark_photon_response}
\end{equation}
or approximately $0.29\%$ of the uncoupled second moment. The two seed
pairs give consistent positive responses. The increase is dominated by
the centroid contribution, $\Delta\langle\bar q^2\rangle=0.0661$~a.u.,
with a smaller internal contribution of $0.00693$~a.u. The positive
second-moment response persists for block lengths from $100$ to
$1000$~fs. A paired block bootstrap that recenters each condition also
gives a positive centered-variance increase, showing that the response
primarily reflects changed fluctuations.

The molecular structural response is much smaller.
Fig.~\ref{fig:benchmark_paired_response}-b gives
$\Delta\langle r_{\mathrm{OH}}\rangle=-2.71\times10^{-5}$~\AA{},
with a $95\%$ confidence interval of
$[-9.97\times10^{-5},\,4.56\times10^{-5}]$~\AA{}.
The H--O--H angle change likewise remains unresolved, with an estimate
of $0.00666^{\circ}$ and a confidence interval of
$[-0.01344^{\circ},\,0.02676^{\circ}]$. Thus, the sampled photon
response is resolved without a statistically resolved change in these
mean molecular geometry measures.

Taken together, the energy-conservation, free-photon, and coupled
sampling benchmarks establish complementary numerical evidence for the
released workflow. The coupled response is a reproducible result of the
specified finite-basis CBO-HF model at $P_{\mathrm b}=32$ and
$\Delta t=0.1$~fs. Users are strongly encouraged to test bead and timestep convergence before interpreting any emerging result, case by case. Finally, comparing the two couplings at fixed quantum discretization does not isolate a nuclear-only or photon-only quantum contribution.

\section{Post-processing utilities}\label{sec:post-proc}
\subsection{Dipole and photon-coordinate spectra}
The analysis scripts provided in this release process dipole and photon trajectories, compare or average spectra, and summarize peaks. The multi-trajectory classical IR utility pools dipole-derivative correlations before Fourier transformation. For laboratory polarization $\mathbf u$, the observable has the form
\begin{equation}
 I_{\mathbf u}(\Omega)\propto\operatorname{Re}
 \int_0^\infty\dd t\,e^{i\Omega t}
 \left\langle[\mathbf u\cdot\dot{\bm\mu}(0)]
 [\mathbf u\cdot\dot{\bm\mu}(t)]\right\rangle_{\mathrm{cl}},
 \label{eq:spectrum}
\end{equation}
with prefactor and normalization chosen for the intended comparison. The photon-coordinate spectrum characterizes $q$ correlations.

The IR utility provides scalar/Cartesian dipole input, five-point differentiation, laboratory or water-fixed frames, finite correlation windows, artificial linewidth, zero padding, peak tables, and trajectory-bootstrap confidence bands. It applies no quantum correction factor. Its charged-trajectory analysis uses $\bm\mu_{\mathrm{int}}=\bm\mu_{\mathrm{lab}}-Q\mathbf R_{\mathrm{COM}}$; this analysis option does not validate charged electronic cavity calculations. Nuclear center-of-mass and frame operations exclude \code{L}.

Reliable spectra require suitable real-time trajectories and a correct sampling interval. Body-frame and laboratory-frame spectra typically reproduce different effects under fixed cavity polarization. The finite correlation time and window determine resolution whilst zero padding interpolates without adding independent information. 

\subsection{Release identity and matched parameters}
The repository can be found at \url{https://github.com/gcassone-cnr/cavity-qed-md}. The exact \code{eT} source archive supplied with this release is identified by its SHA-256 checksum, allowing users to verify its identity and integrity. The installer builds libcint,\cite{libcint2015} applies the gradient patch, runs \code{eT}'s Fortran-source generation, compiles \code{eT}, and stages the workflow. The environment pins \code{i-PI} 3.2.0, NumPy 2.2.6, and SciPy 1.15.3, with recorded analysis/build dependencies. Optional LibXC supports ordinary DFT single points, not yet validated CBO-DFT molecular dynamics. Two example trees retain different defaults, as  listed in Table~\ref{tab:defaults}.  
\begin{table}[tb]
\centering
\caption{Default parameters of the distributed examples. Angular frequencies correspond to approximately 4115.3 and 3656.4~cm$^{-1}$.}
\label{tab:defaults}
\small
\begin{tabularx}{\linewidth}{@{}lXX@{}}
\toprule
Setting & Workstation workflow & Preserved monomer/PIMD \\
\midrule
Polarization & Laboratory $z$ & Laboratory $y$ \\
$\omega$ (a.u.) & 0.0187505 & 0.01666 \\
$\lambda$ (a.u.) & 0.01 & 0.005 \\
Basis & cc-pVDZ, tight SCF & aug-cc-pVDZ \\
Beads & One in supplied MD preparation & Eight in PIMD input \\
Examples & Monomer/cluster NVT/NVE; global Eckart NVE
 & Monomer NVE and joint PIMD \\
\bottomrule
\end{tabularx}
\end{table}

\section{Conclusions}
\code{Cavity-QED Molecular Dynamics} connects \code{eT} and \code{i-PI} to perform atomistic simulations of finite molecular systems coupled to a single photon mode on a self-consistent CBO-HF energy surface. The implementation supports classical nuclear-photon dynamics in the NVT and NVE ensembles, including an optional global Eckart constraint, and canonical joint nuclear-photon path-integral sampling. The included gradient correction restores the contributions required at fixed photon displacement, providing nuclear and photon forces consistent with the same energy surface. OpenMP threading, concurrent bead evaluations, SCF restart recovery, and analysis utilities support practical and reproducible calculations.

The water-monomer benchmarks test both force-driven propagation and equilibrium sampling. Energy-derivative checks and the approximately quadratic timestep dependence of NVE energy fluctuations support the consistency of the electronic energies, forces, and integration scheme. Joint PIMD simulations with PILE-L thermostatting reproduce the analytic finite-bead photon second moment and its centroid and internal contributions within the estimated sampling uncertainties. At the tested parameters, cavity coupling produces a statistically resolved increase in photon fluctuations, while changes in the mean molecular geometry remain a topic for future investigations. These results demonstrate the workflow's ability to resolve a small coupling-dependent equilibrium response at a specified quantum discretization.

By combining self-consistent electronic polarization, analytic forces, and bead-resolved joint path-integral sampling, the released code provides a reproducible framework for investigating how coupled quantum fluctuations of matter and the electromagnetic field shape molecular structure and equilibrium properties within the CBO-HF approximation, paving the way for studies of polaritonic phenomena. Future applications could address cavity-modified hydrogen-bond networks and vibrational spectra in molecular clusters, exploring how nuclear and photonic quantum fluctuations jointly influence these properties in regimes of vibrational strong and ultrastrong coupling.

\section*{Code and data availability}
A public repository providing the source and all needed instructions are available on GitHub at https://github.com/gcassone-cnr/cavity-qed-md 


\section*{Use of AI tools}
OpenAI ChatGPT (Codex) assisted with software development and debugging. AI tools are not authors. Responsibility for scientific content, authorship, and submission rests with the human author.

\end{document}